# Polarization-Stable Long-Distance Interference of Independent Photons for Quantum Communications

T. Ferreira da Silva[1,2*], D. Vitoreti[1], G. B. Xavier[3,4], G. P. Temporão[1], and J. P. von der Weid[1]

[1]*Center for Telecommunication Studies, Pontifical Catholic University of Rio de Janeiro,*
*R. Marquês de São Vicente 225 Gávea, Rio de Janeiro – Brazil*
[2]*Optical Metrology Division, National Institute of Metrology, Quality and Technology,*
*Av. Nossa Sra. das Graças 50, Duque de Caxias, RJ – Brazil*
[3]*Departamento de Ingeniería Eléctrica, Universidad de Concepción, Casilla 160-C, Correo 3, Concepción – Chile*
[4]*Center for Optics and Photonics, Universidad de Concepción, Casilla 4016, Concepción – Chile*
*Author e-mail address: thiago@opto.cetuc.puc-rio.br*

**Abstract:** Interference between fully-independent faint laser sources over two 8.5-km full polarization-controlled fiber links was performed, with stable visibility of 47.8%, an essential step towards practical implementation of quantum communication protocols.

## 1. Introduction

Quantum communication allows the transfer of a quantum state between two remote sites [1]. The maximum achievable distance between two stations, as well as the maximum key generation rate, are severely limited by channel loss, also tailoring a commitment with maximum transmission rate. To overcome this drawback, some quantum repeater protocols were proposed, based on entanglement swapping with quantum memories or linear optics with atomic ensembles [2]. Both methods are technologically challenging, because whereas the first one requires light storage in a quantum memory, in the second one sub-wavelength phase stabilization [3] between two independent fibers, linking the two transmission and the repeater stations, is needed. A third approach entangles two remote atomic ensembles in a mid-way station, which performs a projective measurement of the single photons emitted by the devices, which become, in turn, entangled to each other [4]. The two-photon interference requires them to be indistinguishable at the repeater station, where a Bell-states measurement is performed. The success of this projective measurement requires the arrival time mismatch to be smaller than the coherence time of the photons, assuring overlap of the wave packets [4]. If two photons are made indistinguishable (in polarization, frequency and overlapping of the wave packets), they can interfere in a beamsplitter. This causes the photon bunching effect, in which two photons entering different ports of an optical beamsplitter exit randomly at a common spatial mode. This phenomenon is quite known with entangled pairs, which can achieve unitary visibility. However, when using two Poissonian sources, the maximum visibility is 0.5, due to multi-photon events [5, 6].

In this work we perform a stable long-distance interference measurement between fully-independent faint lasers over two independent spans of full-polarization controlled fiber links, with 8.5 km each. Results demonstrate the feasibility of a controlled channel with current optical fiber technology to be used for quantum communications.

## 2. Experimental setup

A long-distance interference measurement setup with different faint lasers was assembled as depicted in Fig. 2. Two independent laser diodes, with 800 kHz and 6 MHz long term linewidths, were used at Alice and Bob to send polarization states to a middle-way Charlie remote station. Each external cavity continuous wave (CW) laser diode ($L_{Alice}$ and $L_{Bob}$), succeeded by a polarization controller (PC) and a variable optical attenuator (VOA), is tuned to match the 1546.12 nm DWDM channel, corresponding to the quantum channel. The state of polarization (SOP) of the lasers is adjusted to overlap and are sent over optical fiber to reach Charlie's station, entering at each input port of a 50 % beamsplitter (BS). At each output port of thee BS, a single-photon avalanche detector ($SPD_1$ and $SPD_2$) is placed, with an additional hundred meters-long optical delay line at the $SPD_2$ branch. Given that a counting event is recorded at $SPD_1$, it triggers $SPD_2$ after a controlled delay which allows the detector gates to be scanned for coincidence detection at the beamsplitter. The ratio between $SPD_2$ and $SPD_1$ output pulses independently summed over a time interval gives the coincidence count rate.

The automatic full polarization control used in each 8.5-km fiber is similar to the one used for polarization-encoded quantum communications [7], using two ASE-filtered CW telecom lasers emitting in adjacent DWDM channels, respective to the quantum one ($\lambda_{P1}$ at 1545.32 nm and $\lambda_{P2}$ at 1546.92 nm). For sake of simplicity, we share both control lasers between Alice and Bob.

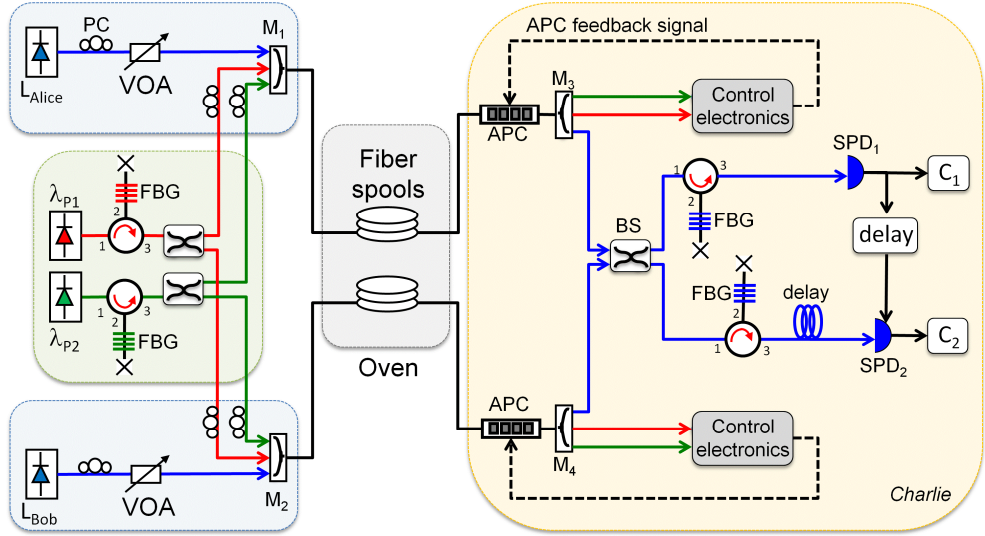

*Figure 1 – Setup for the long-distance interference between independent lasers with full polarization control.*

## 3. Results

The interference dip was measured by scanning the gate delay of $SPD_2$ relative to $SPD_1$, exceeding the photons coherence times. The result is a convolution of the detectors gates and coherence times of the photons. The bunching visibility, defined as $V = (C_{dist} - C_{ind})/C_{dist}$, where $C_{dist}$ and $C_{ind}$ refer to the coincidence ratio with the lasers distinguishable and indistinguishable, respectively, was measured by comparing the coincidence counts with the gates matched and delayed by 1 μs (much longer than the photons coherence times). The dip width was found to depend on the linewidth of the laser sources, as expected, changing from 45 ns down to 11 ns as the wider linewidth was further broadened up to 100 MHz by means of FM modulation. In full implementation of quantum communication protocols, the pulsed lasers will define the coherence time of the photons, which must be longer than the detector gate to guarantee indistinguishability.

Figure 2a shows the behavior of the interference visibility, as the relative SOP of the lasers was scanned from parallel to orthogonal The interference visibility is strongly dependent on the optical intensity ratio of the lasers (R) and, for weak coherent states, follows [6] $V = 2R/(R+1)^2$. This influence was verified by varying the optical power of one laser with the other fixed, emitting 1 photon per detector gate on average. Figure 2b exhibits the measured visibility and the expected values.

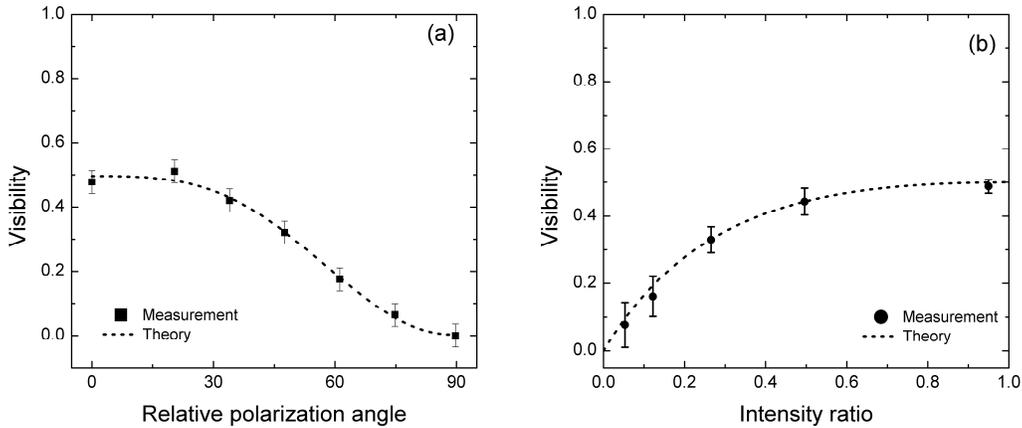

*Figure 2 –Interference visibility as function of (a) polarization mismatch and (b) optical intensity ratio, ranging from 0 to 0.5 for Poissonian sources.*

After making both lasers indistinguishable in polarization and frequency, the visibility stability was measured over time. The temperature of the independent fibers was forced to randomly vary while the coincident count ratio between detectors was measured with gate delay matched and detuned 1 μs. Stable mean visibility of 47.8 % was observed with polarization control on, as seen in figure 3.

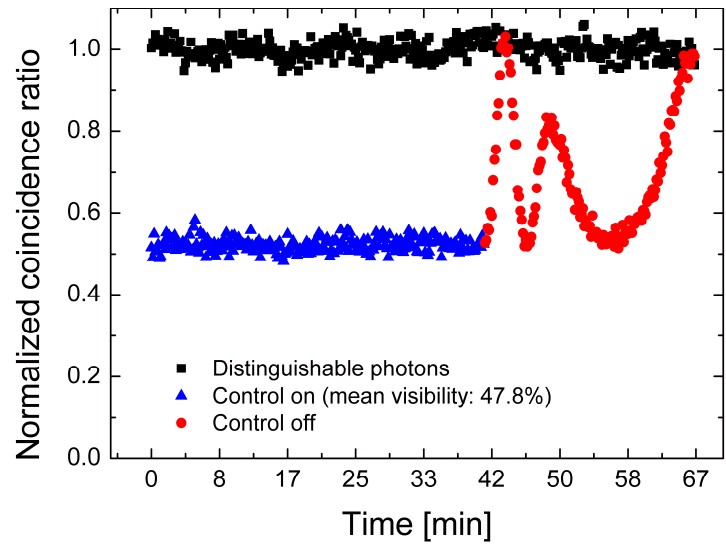

*Figure 3 –Normalized coincidence ratio with polarization control on (triangles) and off (circles). Distinguishable was achieved in both situations by detuning the relative gate delay (squares).*

After a 42 minutes time interval, polarization control was turned off. The fibers birefringence varied randomly and the visibility of interference wandered between minimum (0) and maximum values (0.5), due to the SOP overlap degradation.

### 4. Conclusion

We demonstrated stable interference between photons from fully-independent faint laser sources. Quality of interference was measured when transmitting each laser over an 8.5 km full polarization-controlled fiber spool, with stable 47.8% visibility observed during more than 40 minutes. These results are essential steps towards practical implementation of some quantum communication protocols over optical fiber, like quantum repeaters.

### 5. Acknowledgments

This work is supported by Brazilian agencies CAPES, CNPq and FAPERJ. In addition G. B. Xavier acknowledges support from grants Milenio P10-030-F and CONICYTPFB08-024.